\documentclass[12pt]{article}
\usepackage{graphicx}
\usepackage{url}
\begin{document}
\begin{center}
     {\large\bf Experiment and theory: the case of the Doppler effect for photons }\\
    \vskip5mm
  \small  {Giuseppe Giuliani}\\
    \vskip3mm
 {Dipartimento di Fisica, Universit\`a degli Studi di Pavia, via Bassi 6, 27100 Pavia, Italy}\\
{giuseppe.giuliani@unipv.it}   
\end{center}
\noindent
{\bf Abstract}.
In 1907, Einstein suggested an experiment with flying atoms for corroborating time dilation. In that paper, the flying atom was conceived as a flying clock: the reference to the Doppler effect was only indirect (the experiments by Stark to the first order of $v/c$). In 1922, Schr\"odinger showed that the emission of a light quantum by a (flying) atom is regulated by the conservation laws of energy and linear momentum. Therefore, the Doppler effect for photons is the consequence of the energy and momentum exchange between the atom and the photon:  a central role is played by the quantum energy jump $\Delta E$ of the transition (a relativistic invariant). The first realization of the experiment devised by Einstein is due to Ives and Stilwell (1938). Since then  till nowadays experiments of this kind have been repeated in search of better precision and/or a deviation from the predictions of special relativity. The striking feature is that all the papers dealing with these experiments completely neglect Schr\"odinger's dynamical treatment. The origins of this omission are of different kind: pragmatic (agreement between formulas, wherever coming from, and experiments),  historical (deep rooting of the wave theory of light) and epistemological (neglect of basic epistemological rules).
\vskip5mm\par\noindent
pacs {03.30.+p 01.65.+g}

\section{Introduction}
If light is treated as an electromagnetic wave in vacuum, it is straightforward to derive the formula for the Doppler effect by using, for instance, the transformation equations of the  four~-~wavevector $(\omega/c, k_i)$:
\begin{equation}\label{omegaf}
\omega = \omega\,' {{\sqrt{1-B^2}}\over{1-B\cos \theta}}; \qquad B=\frac{v}{c}
\end{equation}
As usual, we  are dealing with two inertial frames ($O,O'$) whose axis are parallel and oriented along the same directions: $O'$ is considered in motion with respect to $O$ with velocity $v$ along the positive direction of the common $x\equiv x'$ axis; $\theta$ is the angle that the wavevector $\vec k$ forms with the $x$ axis.
\par
If we introduce the concept of photon and write for its energy $E_p=\hbar \omega$, equation (\ref{omegaf}) becomes valid also for the photon energy:
\begin{equation}\label{pho}
    E_p = E_p' {{\sqrt{1-B^2}}\over{1-B\cos \theta}}; \qquad B=\frac{v}{c}
   \end{equation}
Alternatively,  equation (\ref{pho}) can be derived by treating photons as relativistic particles whose energy~-~momentum four~-~vector is $(E/c,\vec p)$, with $E=\hbar \omega$ and $p=\hbar\omega/c$.
\par
     The light source  and its physical state do not enter equations (\ref{omegaf}, \ref{pho}). Therefore, the interpretation  of equation (\ref{omegaf})   reads: if $\omega '$   is the angular frequency of the light wave measured in the reference system  $O'$, then the angular frequency of the same wave  measured by $O$ is given by equation (\ref{omegaf}).  A similar interpretation holds for equation (\ref{pho}). Notice that the source needs not to be at rest  in $O'$.
   \par
  Equations (\ref{omegaf}, \ref{pho}), are used  for describing the Doppler effect also  when experiments deal with photons emitted or absorbed by atoms/nuclei. Moreover,
              since around 1905 till nowadays, the (transverse) Doppler effect for light has been interpreted as an experimental corroboration of time dilation both in research papers and textbooks.
  \par
This paper  discusses these issues in the frameworks of the wave and the corpuscular descriptions of light; it  discusses also  some historical passages bringing out the underlying  epistemological aspects. Therefore, it may be of some interest  for researchers and for teachers at university or high school levels.
\section{Doppler effect for photons\label{dopplersec}}
\subsection{Stark and Einstein around 1905: atoms as clocks}
   Within the theoretical framework around 1900, the emission of light by matter was due to the { harmonic vibrations of  point electrical charges (electrons).  On the basis of this  model, Stark concluded that lines emitted by atoms in flight should have been Doppler shifted. In 1905, he succeeded in detecting this shift (to the first order in $v/c$) for light emitted by hydrogen atoms \cite{stark1}. This achievement, together with the later discovery of the Stark effect, earned him the Nobel prize in physics in 1919 \footnote{Stark discovery of the effect holding his name has been the result of a research program. As discussed in \cite{matteo}, the contemporaneous and independent discovery of the same effect by Lo Surdo, was an accidental one.}.
  \par
Einstein commented the results obtained by Stark with these words:
   \begin{quote}
    In an important paper published last year, Mr. J. Stark ({J. Stark, {\em
Ann. d. Phys.} 21 (1906), 401}) demonstrated that the moving positive ions of
canal rays emit line spectra by identifying the Doppler effect and following
it quantitatively. He also undertook experiments with the intention of
detecting and measuring an effect of the second order (proportional to
$(v/c)^2$); however, the experimental arrangement, which was not set up specifically
for this purpose, was not adequate for achieving reliable results.
\par
I will show here briefly that the principle of relativity
in conjunction with the principle of the constancy of the velocity of light
makes it possible to predict the above effect.
As I showed in an earlier paper, it follows from these principles that
a uniformly moving clock runs at a slower rate as judged from a
``stationary'' system than as judged by a co~-~moving observer.
 If $\nu$ denotes the number of the clock's strokes per unit time for
 the observer at rest, and $\nu_0$ the corresponding number for the
 co~-~moving observer, then
 \begin{equation}\label{fre}
    {{\nu}\over{\nu_0}}  =\sqrt { 1 - \left[{{v}\over{c}}\right]^2  }
\end{equation}
or to first approximation
\begin{equation}\label{freappr}
 {{\nu-\nu_0}\over{\nu_0}}  = -{{1}\over{2}} \left[{{v}\over{c}}\right]^2
\end{equation}
The atom ion of the canal rays that emits and absorbs radiation of certain
frequencies is thus to be conceived as a fast~-~moving clock, and the
relation just indicated can therefore be applied to
 it \cite[p. 232]{ein07}.
   \end{quote}
Both Stark and Einstein assumed that the {\em atom is a clock}. Clearly, this assumption is justified only if the atom is the seat of some periodic motion. This motion could be, for instance, the supposed harmonic motion of  bound electrons; or the supposed elliptical motion of the electron in Bohr's model of hydrogen atom: however, in this case, the frequency of the light emitted by the hydrogen atom differs from the frequency of the electronic motion. Anyway, the advent of quantum mechanics forbids any description of  atoms as seats of periodic motion of electrons: {\em atoms are not clocks} \footnote{The fact that we now use {\em atomic} clocks should not confuse us: in these clocks, a quantum transition between two atomic levels is used as a locking parameter of the resonant frequency of a quartz oscillator.}.
\par
Nevertheless, let us come back to Einstein.
 After having assumed that
 atoms are clocks,  he
applied to them the time dilation formula. He thus obtained equation (\ref{fre}) that yields the ratio between the ``proper'' frequency $\nu_0$ of an atom~-~clock and the frequency $\nu$ of the same atom~-~clock for an observer that sees the atom~-~clock in uniform motion with velocity $v$.
\par
Einstein did not refer to  the formula  for the Doppler effect (\ref{omegaf})  derived  two years before in his theory of special relativity  by treating light as an electromagnetic wave \cite[p. 161]{ein05}. In this paper, Einstein proved also that the energy of a ``light complex'', i.e. the  light energy contained in a sphere moving at the speed of light with respect to an inertial frame, in passing from this frame to another, changes in exactly the same way as the light frequency. Of course, this means that the light energy and its frequency are connected by a relativistic invariant: $U = (Nh)\nu$. However, Einstein did not take this step: as stressed and commented by Pais, Einstein kept well separated special relativity from the light quanta hypothesis \cite[p. 909]{pais}.
\subsection{Schr\"odinger, 1922: atoms emit light quanta endowed with energy and momentum \label{erwinsec}}
In 1922 Schr\"odinger dealt with the radiation emitted
   by atoms in motion in terms of light
quanta \cite{erwin}.
Schr\"odinger makes it clear from the beginning that, once we accept Einstein's idea
that the quantum $h\nu$ `always carries' a linear momentum $h\nu/c$, we have to
recognize that the emission of a quantum $h\nu$ by an atom produces a ``jump''
in its velocity and that {\em this jump is responsible
for the Doppler shift}.
\par
The problem posed by Schr\"odinger is illustrated in fig. \ref{erwinfig} (not used by Schr\"odinger).
In the reference frame of the measuring apparatus, it is solved by writing down the conservation equations for  energy:
\begin{equation}\label{energia2}
E_p=\gamma_1 E_1-\gamma_2 E_2
\end{equation}
and momentum:
\begin{eqnarray}
\gamma_1 {{E_1}\over{c^2}}v_1 \cos\theta_1 &= & \gamma_2
{{E_2}\over{c^2}}v_2 \cos\theta_2 +{{E_p}\over{c}}
\label{qmx}\\
\gamma_1 {{E_1}\over{c^2}}v_1 \sin\theta_1 &= & \gamma_2
{{E_2}\over{c^2}}v_2 \sin\theta_2\label{qmy}
\end{eqnarray}
\begin{figure}[htb]
 \centerline{
 \includegraphics[width=5cm]{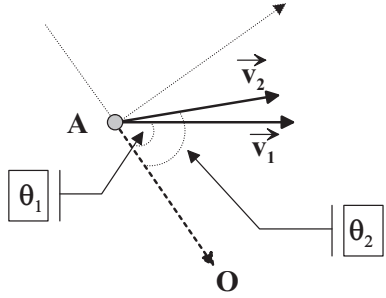}
 }
  \caption{\label{erwinfig}
  Emission of a light quantum  by the atom $A$ in  motion. The light quantum is emitted along the direction $A\rightarrow
O$: $O$ is the entrance slit of the spectrograph. The subscript $1$ denotes the quantities  before the emission;  the subscript $2$ the quantities after the emission.
  }
 \end{figure}
\par\noindent
     $E_p$ is the energy of the light quantum; $E_1$ and $E_2$  are the rest energies of the atom before and after the emission, respectively;  $\gamma_1,\,\gamma_2$ are the relativistic factors before and after the emission.
\par
After a somewhat lengthy analytical manipulation we get:
\begin{equation}\label{uguale}
E_p =E^0_p  {{\sqrt{1-v_1^2/c^2}}\over{1 -(v_1/c)\cos \theta_1 }}
\end{equation}
with
\begin{equation}\label{zero}
 E^0_p= \Delta E \left( 1 - {{\Delta
E}\over{2E_1}}\right),\: (v_1=0)
\end{equation}
where $\Delta E= (E_1-E_2)$; then,  $\Delta E$  is  the energy difference between the two states of the atomic transition. $E^0_p$ is the measured light quantum energy when the atom is at rest before the  emission. {\em Both $\Delta E$ and $E_p^0$ are relativistic invariants since they depend only on rest energies}. The term $\Delta E/2E_1$ is in general negligible, unless we are dealing with $\gamma$ photons emitted by free nuclei.
\par
Schr\"odinger's did not write equation (\ref{uguale}). He stopped at an intermediate formula containing both atoms' velocities, before and after the emission \cite[p. 303]{erwin}:
  \begin{equation}\label{finale2}
  \nu = \nu^* {{1}\over
 {\sqrt{\gamma_1 \left[1 - (v_1/c) \cos \theta_1 \right]\times
\gamma_2 \left[1 - (v_2/c) \cos \theta_2 \right]  }}}
\end{equation}
where
\begin{equation}\label{fstar}
\nu^* =  { {E_1^2 -E_2^2}\over {2h\sqrt{E_1E_2}}  }
 \end{equation}
  A simple calculation suggested by the fact that the atom's velocity  after the emission is determined by its velocity before the emission and by the   energy~-~momentum conservation leads to the more significant equations (\ref{uguale}) and (\ref{zero}) \cite[p. 197-203]{erwingg}. Obviously, Schr\"odinger was well aware of the relation between the two velocities; this renders more intriguing the fact that he stopped at equation (\ref{finale2}).
\par
 Schr\"odinger's treatment can be  applied
also to the case of a photon absorbed by an atom in flight with respect to the
laboratory reference frame \cite[p. 201-202]{erwingg}. The energy $E_p$ that a photon must have for being absorbed by the atom
 is again given by equation (\ref{uguale}), where, in this case:
\begin{equation}\label{nu03}
E_p^0 = \Delta E\left( 1 + {{\Delta E}\over{2E_1}}\right)
\end{equation}
is the energy of the photon  absorbed by an atom at rest
before the absorption and, of course, now $\Delta E=E_2-E_1$.
\par
          Equation  (\ref{uguale}) is formally identical to equation (\ref{pho}) or, since $E_p=\hbar \omega$, to equation (\ref{omegaf}): however, the  meanings of these  formulas are quite different. While equations (\ref{omegaf}, \ref{pho}) ignore the emitting/absorbing particle and its physical state,  equation (\ref{uguale}), through equation (\ref{zero}) for emission or equation (\ref{nu03}) for absorption, focuses on the energy difference $\Delta E$   between the two quantum levels of the emitting/absorbing particle and the energy $E_p^0$ emitted/absorbed when the particle is at rest before emission/absorption (equations \ref{zero}, \ref{nu03}).  Furthermore: while equations (\ref{omegaf}, \ref{pho}) connect quantities measured in two reference frames, Schr\"odinger's treatment uses only the reference frame of the experimental apparatus owing to the use of the relativistic invariants $\Delta E$ and $E_p^0$.
          \par
                    In order to further clarify the physical meaning of  Schr\"odinger's treatment, let us  illustrate the features of the emission/absorption process as described by equations (\ref{uguale}) and (\ref{zero}, \ref{nu03}).
    In the case of photons emitted/absorbed,
    the increase/decrease of their energy with respect to the quantity $\Delta E$ is due to an energy~-~momentum exchange with the emitting/absorbing particle. For instance, let us consider an atom that emits a photon. If the photon is emitted in the forward direction, its energy is increased, with respect to $\Delta E$, by exactly the same amount by which the kinetic energy of the atom is decreased; if the photon is emitted in the backward direction, its energy is decreased, with respect to $\Delta E$, by exactly the same amount by which the kinetic energy of the atom is increased. In the case of absorption, the photon energy required for exciting the atom, when the photon flights against the atom, is decreased,  with respect to $\Delta E$, by exactly the same amount by which the kinetic energy of the atom is decreased; when the photon is chasing the atom, the photon energy required for exciting the atom, is increased,  with respect to $\Delta E$, by exactly the same amount by which the kinetic energy of the atom is increased.
    \par
                  Nowadays, this energy--momentum exchange is basically exploited   by those who laser--cool the atoms or use saturation spectroscopy.
                  \par
                  Schr\"odinger's approach can be  generalized by taking into account the dependence of the energy of the photon emitted/absorbed on the gravitational potential. As shown, for instance by M{\o}ller \cite[pp. 401~-~407]{moller}, it is sufficient to rewrite, for the emission case, equation (\ref{zero}) as follows:
  \begin{equation}\label{nu+-approx}
E^0_p \approx \Delta E\left( 1 - {{\Delta E}\over{2E_1}}\right)
 \left(1+ {{\phi}\over{c^2}} \right)
\end{equation}
where $\phi$ is the gravitational potential and the $\approx$ sign is due to the approximation for  small gravitational potential.
   An analogous correction must be made for equation (\ref{nu03}) (absorption case).
      \par
      Schr\"odinger's paper has been rapidly forgotten; its derivation has been rediscovered by Davisson \cite{dav};   more recent ones can be found, for instance, in the books by French \cite[pp. 197~-~199]{french} and M{\o}ller \cite[pp. 401~-~407]{moller}. No one quotes Schr\"odinger's paper.
      The idea of describing the interaction of photons with particles by writing down the conservation equations  for energy and linear momentum was used some years later by Compton \cite{compton} and Debye \cite{debye}, without quoting  Schr\"odinger, for explaining the Compton effect.
   The reasons why  Schr\"odinger's paper has been forgotten are not clear: we can only guess some plausible ones. First of all, around 1920, the concept of light quanta had not yet been accepted by the scientific community; secondly, the fact that Schr\"odinger stopped at equations (\ref{finale2}) and (\ref{fstar}) might have obscured the relevance of the paper.

    \subsection{Ives and Stilwell, 1938: atoms as clocks, again\label{ivessec}}
    In late Thirties of past century, Ives and Stilwell set up an  apparatus for the
realization of the experiment devised by Einstein in 1907 \cite{ives}.
 The experiment was fully financed
by  the Bell Telephone Laboratories where Ives was then
 working.
 It was a weird twist of fate that
a test of relativity suggested by its founder has been taken up,
performed and interpreted by an anti~-~relativist.
\begin{quote}\small
  Ives (1882~-~1953) has been a staunch opponent
of relativity. He
  builded his views, characterized by an ether~-~based theory,  through
 a series of papers written in about fifteen years: they have been collected in a
  volume \cite{iveslibro}. Unfortunately,  the grossly anti~-~relativist preface by one of the editors throws  an unfavorable  shadow on Ives' papers. A review of this book by Arthur Miller can be found in \cite{ivesrev}.
  Ives' theory is a very intricate one and lies on  a  procedure for  clocks synchronization based on the use of two pairs of rods and clocks, one pair being, by assumption, not affected by their motion in the ether. By assumption,
 the velocity of light is isotropic and equal to $c$ only in
  a reference frame at rest in the ether.
  Ives' coordinates transformations converge to Lorentz's as the velocities of rods and clocks used for synchronization  go to zero. Thus, in principle, the two coordinate transformations could  never coincide. According to Miller, ``Ives's Larmor~-~Lorentz theory was never developed to the point where it could
be seriously considered as an alternative to the special relativity theory''.
 \end{quote}
  The article's title is unequivocal: ``An experimental study of the rate of a moving atomic clock'. As in Einstein's paper \cite{ein07}, the atom is considered as a clock. Ives and Stilwell observe that, as far as the transverse Doppler effect is concerned
   \begin{quote}
    \dots it would be extremely difficult to be sure that observation was made exactly
at right angles to the direction of the rays, and very small deviations from this
direction would introduce shifts of the order of magnitude of
 the expected effect \cite[p. 215]{ives}.
 \end{quote}
 However, this difficulty
\begin{quote}
\dots can be avoided by observing not at right angles, but in two directions, with
and against the motion of the particles; the observation being made simultaneously
by the use of a mirror in the tube. Under these conditions the displaced Doppler lines
are observed corresponding to motion toward and away from the observer, and the
effect to be observed is a shift of the center of gravity of the displaced lines with
respect to the undisplaced line.  As shown in an earlier paper of this series this
shift of center of gravity is expressed by the equation $\lambda =\lambda_0 (1-v^2/c^2)
^{1/2}$ where $v$ is the observed or measured velocity of the positive
 particles \cite[p. 216]{ives}.
\end{quote}
  Since the  experimental setup used by Ives and Stilwell has inspired many experimenters up to nowadays, it is worth discussing it in some details: see the Appendix.
  The experiment allowed to measure the transverse Doppler effect in the approximation of small velocities:  of course, in the  conclusions, Ives and Stilwell interpret their
results as a confirmation of Ives's ether based theory.
  \par
An year later,  Robert Clark Jones, he too at  Bell Laboratories, interpreted Ives and Stilwell
results in a special relativity approach, assuming, of course, that
atoms can be treated as clocks.
Jones did not attack  Ives' standpoint; he wrote instead:
\begin{quote}
The conceptual background of these theories [Larmor's and Lorentz's] is not the
one which is most popular with physicists today, however, and for this reason
it seemed worth while to obtain the theoretical predictions from the point of
view of the special theory of relativity, particularly {\em since the relativistic point
of view yields the results in so simple a manner}. The theoretical
predictions we shall obtain here are identical with those obtained by Ives from electron
theory \cite[p. 337]{jones} (my italics).
\end{quote}
Some years later, a similar experiment with canal rays has been carried out by Gerhard
Otting \cite{otting}.
   The contrast with Ives  and Stilwell's  paper
is striking. Otting does not comment
  either the formulas or their interpretation: he  is interested only
 in  the  correspondence between formulas and experimental data.
\par
The experiment has been repeated again in the sixties by Mandelberg and
 Witten \cite{mand}.
 The authors recall that
 \begin{quote}
  An analysis of  the experiments of Ives and Stilwell and of Otting indicates that although their
reported experimental points seem to fit the curve with
an accuracy of about $2$ to $3\%$, the experimental uncertainty
is more nearly $10~-~15\%$ \cite[p. 529]{mand}.
 \end{quote}
 Hence, the necessity of repeating the experiment.
The basic experimental setup was the same as that of the older
experiment: only the precision was improved. According to Mandelberg and Witten ``The experimental
result is that the exponent in the quadratic
expression for the Doppler shift, $(1-B^2)^{1/2}$, is found to
be $0.498\pm 0.025$ \cite[p. 529]{mand}.'' ``This implies an over-all precision in this experiment of $5\%$ the limit on the accuracy being imposed by the width of the beam lines \cite[p. 536]{mand}.''
 \section{Doppler effect for photons as a {\em direct} consequence of Lorentz transformations\label{dopplerlorsec}}
 Starting about 1970,
   the Doppler shift of radiation emitted/absorbed by atoms/nuclei in flight has been  viewed as a direct consequence of Lorentz transformations: the idea that the atoms are clocks has gone, at least as an explicit statement.
            \subsection{Experiments with $\gamma$ photons}
 Olin et al.  used
 an experimental set up similar to that of Ives and Stilwell \cite{olin}:
 the Doppler shift of  $8.64\, MeV\,\gamma$ photons  emitted by $^{20}Ne$ nuclei was studied. The velocity of the emitting nuclei was $0.012\, c$ or $0.049\, c$. The detector was an annular $Ge(Li)$ junction  that measures the energy of $\gamma$ photons.
  In order to test special relativity through the transverse Doppler effect, the authors discuss the formula:
  \begin{equation}\label{olin}
    E(\theta)=E_0\frac{F(\beta)}{1-\beta\cos\theta}
  \end{equation}
  where the significant quantity is the photon energy and the function $F(\beta)$ is equal to $(1-\beta^2)^{1/2}$ if special relativity is correct ($\beta=v/c$).
They write:
 \begin{quote}
    This phenomenon [Doppler shift] is a
geometrical property of space~-~time, and
is intimately connected with the problem of
synchronization of clocks in different frames
of reference \cite[p. 1633]{olin}.
 \end{quote}
 \subsection{Direct observation of the transverse Doppler shift in Hydrogen}
The transverse Doppler shift of the $H_\alpha$ hydrogen line has been observed {\em directly} (i.e. perpendicularly with respect the direction of the atoms' motion) by Hasselkamp et al. \cite{hass}. The detector was a photomultiplier used in the single photon counting mode.
 According to the authors:
 \begin{quote}
    Equation (\ref{omegaf}) is a consequence of the Lorentz transformation of time. The experimental confirmation of the validity of (\ref{omegaf}) is therefore a verification of time dilation \cite[p. 152]{hass}.
 \end{quote}
\subsection{Lasers enter the scene}
While, from Ives and Stilwell's experiment, the Doppler shift has been studied by measuring the energy of the emitted photons, a major change occurred with the appearance  of lasers: the measured quantity became the energy of the photons  absorbed by the atoms in flight.
The measuring techniques varied from the simple use of lasers for exciting a quantum transition of the atoms in flight \cite{mac}, to the utilization of  two photons absorption \cite{kai} or saturation spectroscopy \cite{saat};  collinear (probing laser beam parallel/antiparallel to the atoms' motion),  orthogonal (probing laser beam perpendicular to the atoms' motion) and variable (different angles between the directions of the probing laser beam and the atoms' motion) geometries have been used.
For a rather recent  review, see \cite{gwinner}.
\par
These papers consider the Doppler shift as a direct consequence of Lorentz transformations and compare the predictions of special relativity with those of the Mansouri~-~Sexl  kinematic test theory of special relativity \cite{sexl}.
This theory is based on the assumption  that the speed of light is isotropic only in a hypothetical  preferred reference frame and use  generalized coordinates transformations that take into account also the possibility of different clocks synchronization procedures.
\par
 Very recently, Chou et al. have studied the transverse Doppler  effect and the gravitational red shift using an optical clock \cite{chou}.
 The unique feature of
 optical clocks consists in the use a {\em single}  ion  at rest in an electromagnetic trap. As far as the transverse Doppler shift is concerned,
Chou et al. have compared the frequency of the probing laser when the ion is at rest (with respect to the laboratory) with that of the same probing laser when the ion is set in  a harmonic motion along a direction  approximately perpendicular to the direction of the  laser beam: as in the other experiments with lasers, the experiment checks the absorption of photons by the ion in motion. Chou et al. interpret their results on the basis of the equation:
\begin{equation}\label{chou}
    \frac{\delta f}{f_0}=\frac{1}{<\gamma (1-v_\|/c)>}-1\approx -\frac{1}{2}\frac{<v^2>}{c^2}
\end{equation}
where $v_\|$ denotes the component of the ion speed along the direction of the probing laser beam; equation (\ref{chou}) is, of course, another way of writing equation
 (\ref{omegaf}) by averaging over time.  $<v_\|>=0$ because the ion's motion is harmonic; $1/\gamma\approx -(1/2)<v^2>/c^2$ because  $v\approx 10\, m s ^{-1}\ll c$.
 Since the experimental data fit equation (\ref{chou}), Chou et al. conclude that they have experimentally tested time dilation.
  \section{Experiments, formulas and theories: what are we measuring? \label{what}}
 Sections (\ref{dopplersec}, \ref{dopplerlorsec})  show that all the papers written after Schr\"odinger's seminal article completely ignore its  treatment; instead, they rely on equations (\ref{omegaf}, \ref{pho})  that compare the angular frequencies of  a light wave or the photon energies in two distinct reference frames.  Let us begin with equation (\ref{omegaf}).
 It is well known that we can derive a  relativistic Doppler formula  valid for both acoustic or light signals  \cite{doppleracot, doppleracot2}:
  \begin{equation}\label{dacot}
    \frac{\omega_a}{\omega_e}=\frac{1-(v_a/V)\cos (\vec V, \vec v_a)}{1-(v_e/V)\cos (\vec V, \vec v_e)}   \frac{\sqrt{1-v_e^2/c^2}}{\sqrt{1-v_a^2/c^2}}
\end{equation}
The reference system is the one in which the medium is at rest; $V$ is the signal velocity; $v_e$ and $v_a$ the emitter and the absorber velocity. For light in vacuum, this formula reduces  to  formula (\ref{omegaf}).
  Equation (\ref{dacot}) is obtained by  assuming that either the source emits signals of ideally null duration at a specified time interval or a periodic wave.  In the first case, the phenomenon's period  is the time interval between two consecutive signals; in the latter, it is the wave period.
  \par
  However, atoms and nuclei  do not emit waves or null duration signals, but photons endowed with energy and linear momentum: therefore, equation (\ref{omegaf}) can not be applied to atoms or nuclei \footnote{Photons or even single photons can be used as light signal of ideally null duration; however, in this case, the frequency  measured is that of the signal period, not the one given by the relation $\nu=E_p/h$ which is the object of this paper.}. Nevertheless, equation (\ref{omegaf}) describes the experimental data: this is due to the fact that equation (\ref{uguale})  reduces to (\ref{omegaf}) by assuming $\omega'=E_p^0/\hbar$, where $E_p^0$ is given by equation (\ref{zero}) for emission or by equation (\ref{nu03}) for absorption and that the velocity entering equation (\ref{omegaf}) is the atom/nucleus velocity before emission/absorption.  Then,  equation (\ref{omegaf}) can be used only through a series of conceptual shifts in passing from a  treatment of the emission/absorption process in term of photons, based on relativistic invariants ($\Delta E, E_p^0$) and one reference system (that of the experimental apparatus), to a formula belonging to the wave theory of light and connecting two quantities (the angular frequencies) in two distinct reference frames. The use of equation (\ref{pho})  does not involve the wave theory of light: however, also in this case, the various conceptual and approximation steps  should be made explicit.
  \par
  The experiments discussed in sections (\ref{dopplersec}, \ref{dopplerlorsec}) are easily explained using Sch\"odinger's approach within the reference frame of the experimental apparatus.
  Let us first consider the experiments in which the flying particle emits photons.
In the laboratory reference frame, given an atom/nucleus and the two quantum levels of the transition (i.e. $\Delta E$), equation (\ref{zero}) yields $E_p^0$; alternatively, $E_p^0$ can be  measured directly when the particle, before emission, is at rest. Then equation (\ref{uguale}) predicts the energy of the photon emitted by the flying particle in terms of the measured velocity $v_1$ and  the angle $\theta_1$. If the experimental test is positive, it corroborates a prediction of the joint use of relativistic dynamics and quantum mechanics.
\par
The experiments with lasers deserve a separate discussion because they deal with {\em absorption} of photons by atoms/nuclei.
  In the laboratory reference frame, given an atom/nucleus and the quantum levels of the transition (i.e. $\Delta E$), equation (\ref{nu03}) yields $E_p^0$; alternatively, $E_p^0$ can be  measured directly
 when the particle, before absorption, is at rest. Then, given the laser photons energy $E_p^0$, the particle velocity $v_1$ and the angle $\theta_1$, equation (\ref{uguale}) yields the energy  that the laser photon must have in order to be absorbed by the flying particle. If the experimental test is positive, i.e. if the flying particle absorbs the laser photon, then it corroborates a prediction obtained by the joint use of relativistic dynamics and quantum physics.
 \par
 The measurement of the gravitational red shift by Chou et al. \cite[p. 1632]{chou}, calls for a final remark. The generalization of Schrodinger's treatment that takes into account the gravitational potential yields:
  \begin{equation}\label{dopplergrav}
    E_p\approx \Delta E\left( 1 + {{\Delta E}\over{2E_1}}\right)
 \left(1+ {{\phi}\over{c^2}} \right) {{\sqrt{1-v_1^2/c^2}}\over{1 -(v_1/c)\cos \theta_1 }}
  \end{equation}
  where, as already pointed out in section \ref{erwinsec}, the $\approx$ sign comes in for small gravitational potentials.
  Equation (\ref{dopplergrav}) describes contemporaneously both effects: the  Doppler effect due to the velocity of the absorbing particle and the gravitational one. Instead, in the quoted paper, the two effects seem to derive from two distinct theoretical backgrounds.
  \par
            Why Schr\"odinger treatment keeps being neglected?
            The first and basic answer  is a pragmatic one:  as pointed out, formulas (\ref{omegaf}, \ref{pho}) describe the experimental data. However, physics is not  a game between formulas and experiments: formulas belong to theories with a well defined application domain. Therefore, formulas can not be extrapolated from a theory and applied to phenomena belonging to other theoretical frameworks. As Heinrich Hertz put it some time ago:
            \begin{quote}
            The very fact that different modes of representation contain
what is substantially the same thing, renders the proper
understanding of any one of them all the more difficult. Ideas
and conceptions which are akin and yet different may be
symbolised in the same way in the different modes of representation.
Hence for a proper comprehension of any one of
these, the {\em first essential is that we should endeavour to understand
each representation by itself without introducing into it
the ideas which belong to another} \cite[p. 21]{hertz_ew} (my italics).
   \end{quote}
       Applied to our case, this means that we should not mix ideas and formulas coming from two different ``representations'' as the undulatory and the corpuscular theory of light. We should instead  define their  domains of applications and find out  when, how and why their predictions coincide.
             \par
             But there is another reason:
                      the nineteenth century has deeply embedded the undulatory description of light in the background knowledge of physicists. Within this  knowledge, the Doppler effect has been constantly viewed as a wave phenomenon: the emergence  of the light quantum did not change this rooted habit.
   The prevailing influence of the undulatory description emerges also in the language: for instance, the locution `photon frequency' is often used instead of  `photon energy' and, in general, when dealing with photon the equations are written in terms of frequencies instead of energies.
  \par
  There  are two questions left: a) the Doppler effect as a direct consequence of Lorentz transformations; b) the statement according to which the experimental corroboration of the Doppler effect for photons is a corroboration of time dilation.
  \par
  About a).  As we have seen, the Doppler effect for photons is a consequence of the relativistic conservation laws for energy and momentum and the concept of photon endowed with energy and linear momentum: the term ``direct'' is clearly inappropriate since we must add to Lorentz transformations the laws of relativistic dynamics and the quantum concept of atoms/nuclei energy levels.
    \par
    Instead, b) is a sound statement. With a specification: it is an indirect corroboration. The  relativistic factor $\sqrt{1-v^2/c^2}$,  basically a time dilation factor,  enters equation (\ref{uguale}) through the relativistic conservation equations; therefore, an experimental corroboration of equation (\ref{uguale}) is, primarily, a corroboration of relativistic dynamics and some quantum hypothesis. That the factor  $\sqrt{1-v^2/c^2}$ is a time dilation factor, can be  easily shown by ideal experiments with light signals of null duration. This kind of approach has been firstly outlined by Bondi \cite{bondi}, who, however, uses also geometric considerations.  Derivations of time dilation, length contraction and Lorentz transformations based only on  ideal experiments with light signals of null duration can be found in,  for instance, \cite{uga} and \cite[p. 29-42]{erwingg}.
     \par
    Then, every corroboration of a  formula containing the relativistic factor $\sqrt{1-v^2/c^2}$ can be considered as an indirect corroboration of time dilation.
    For example, a typical formula of this kind is the one giving the radius $R$ of  the circular  trajectory  of a point electrical charge that enters an uniform magnetic field perpendicularly to it.
  With obvious notations, we have (neglecting irradiation):
   \begin{equation}\label{raggio}
    R= \frac{1}{\sqrt{1-v^2/c^2}}\frac{mv}{qB}
 \end{equation}
Since equation (\ref{raggio})  is derived by the joint use of  the relativistic force law  and the expression of Lorentz force, its experimental verification constitutes a corroboration of both laws and, {\em indirectly}, of time dilation.

      \section{Conclusions}
      A survey  of the experiments on the Doppler effect for photons (through about a century) shows that the explanation of these experiments has been initially given by considering the atoms as clocks; then by using the Doppler formula of the wave theory of light (or its  purely formal reformulation in terms of photon energies). The   corpuscular treatment of the Doppler effect put forward by Schr\"odinger in 1922   has been completely ignored in spite of  the fact that, in other physical contexts, it is commonplace that during the emission/absorption of a photon by an atom/nucleus the emitting/absorbing particle exchanges energy and momentum with the photon: for instance, in the emission/arborption of photons by free nuclei, in the laser cooling of atoms and in saturation spectroscopy.
    \par
    The origins of this omission are of different kind: pragmatic (agreement between formulas, wherever coming from, and experiments),  historical (deep rooting of the wave theory of light) and epistemological (neglect of basic epistemological rules).
      \par
      Physics is not simply a game between formulas, from wherever they are, and experiments: one should not use formulas from one theory (the wave theory of light) to describe phenomena concerning photons which belong to another theory (the corpuscular theory of light). This epistemological criterium, put forward in its general form more than a century ago by Heinrich Hertz, should not be forgotten, particularly in teaching; however, also in research, the neglect of basic epistemological criteria may, in the long run, lead astray.
\section{Appendix: Ives and Stilwell experiment}\small
The formulas appearing below from (i) to (iv) are relativistic formulas, not Ives'. However, the central result given by equation (\ref{dt}) is the same also for Ives.
  \begin{enumerate}
 \item The wavelength $\lambda_B$ of the light emitted by the
 incoming atoms
  at a small angle $\theta$ to the beam direction is given by:
 \begin{equation}
\lambda_B=\lambda_0 {{1-B\cos \theta}\over{(1-B^2)^{1/2}}}\approx \lambda_0\left(1-B\cos \theta+
{{1}\over{2}}B^2\right)
\end{equation}
where $\lambda_0$ is the natural wavelength, $B=v/c\ll 1$ and
 $v$ is the velocity of the
atoms.
\item The wavelength $\lambda_R$ of the light emitted by the
 receding atoms at the angle $(\pi-\theta)$ to the beam direction is instead:
\begin{equation}
\lambda_R=\lambda_0 {{1+B\cos \theta}\over{(1-B^2)^{1/2}}}\approx \lambda_0\left(1+B\cos \theta+
{{1}\over{2}}B^2\right)
\end{equation}
 \item The average of the two wavelengths is:
\begin{equation}\label{dt}
\lambda_Q={{1}\over{2}}(\lambda_B+\lambda_R)=
 \lambda_0\left(1+
{{1}\over{2}}B^2\right)
\end{equation}
and equals the wavelength which would be observed at right angles to the beam.
\item The difference between the two wavelengths is:
\begin{equation}\label{lamd}
2\lambda_D=\lambda_R-\lambda_B=2\lambda_0 B \cos \theta
\end{equation}
\end{enumerate}
where $\lambda_D$ represents the first order Doppler effect.
\par
 The emitting particles where  hydrogen atoms. The observed
 line, on a photographic plate, was the  $H_\beta$ line.
Ives and Stilwell:
\begin{description}
\item[a)] derived the experimental value of the atoms velocity $v$ from (\ref{lamd}), using for $\lambda_0$ the wavelength of the ``undisplaced line'', i.e. the central line appearing on the plate together with the ``displaced lines'' $\lambda_B$ and $\lambda_R$;
\item[b)] predicted the value of the transverse Doppler shift by using this value
of the atoms velocity;
\item[c)] compared the calculated value of the transverse Doppler shift with
that measured according to (\ref{dt});
\item[d)] concluded by stating that the theoretical predictions agree with experimental results within
measurements precision \footnote{As a matter of fact, Ives and Stilwell used also an a') step (instead of a)) in which
 the velocity of the emitting atom was assumed to be the velocity of the incoming
 $H_2^+$ or $H_3^+$ ion calculated through the relation $eV=(1/2) M v^2$, where
 $e$ is the charge of the ion, $M$ its mass, $v$ its velocity and $V$ the accelerating
 potential. Ives and Stilwell found that both procedures for calculating the velocity
 of the emitting atoms resulted in a predicted transverse Doppler shift
 (point b) above) in agreement with the measured one (point c) above).}.
\end{description}

\end{document}